# *CPT*-Frames for *PT*-symmetric Hamiltonians

Huai-Xin Cao, Zhi-Hua Guo, Zheng-Li Chen
College of Mathematics and Information Science,
Shaanxi Normal University, Xi'an 710062, China

Email: caohx@snnu.edu.cn

**Abstract:** *PT*-symmetric quantum mechanics is an alternative formulation of quantum mechanics in which the mathematical axiom of Hermiticity (transpose and complex conjugate) is replaced by the physically transparent condition of space-time reflection symmetry (*PT*-symmetry). A Hamiltonian $H$ is said to be *PT*-symmetric if it commutes with the operator *PT*. The key point of *PT*-symmetric quantum theory is to build a new positive definite inner product on the given Hilbert space so that the given Hamiltonian is Hermitian with respect to the new inner product.

The aim of this note is to give further mathematical discussions on this theory. Especially, concepts of *PT*-frames, *CPT*-frames on a Hilbert space and for a Hamiltonian are proposed, their existence and constructions are discussed.

## 1. Introduction

It is well-known that in a classical quantum mechanics system, the time evolution of the system is described by the Schrödinger equation of a Hamiltonian $H$, which is a densely-defined Hermitian operator in the Hilbert space $K = L^2(\mathbf{F}^n)$ (where $\mathbf{F} = \mathbf{R}, \mathbf{C}$). The Hermiticity of $H$ ensures that the spectrum of $H$ is real. It is remarkable that the Hermiticity of a Hamiltonian $H$ is not necessary for the reality of the spectrum. Bender and Boettcher observed in [1] that the reality of spectrum of a Hamiltonian $H$ is due to $PT$ symmetry of $H$. Since *PT*-symmetry theory has many really physical backgrounds and applications, it has been widely discussed and developed [2-16], and many conferences on this theory have been held. For more recent work on this topic, please refer to [17-25].

Mathematically, the parity $P$ and the time reversal $T$ are indeed the operators defined as follows.

$$(Pf)(x) = f(-x), \quad (Tf)(x) = \overline{f(x)}, \quad \forall f \in L^2(\mathbf{F}^n), \tag{1}$$

where the bar means the complex conjugate and $\mathbf{F}$ denotes the real field $\mathbf{R}$, or complex field $\mathbf{C}$. The canonical inner product on the Hilbert space $L^2(\mathbf{F}^n)$ is given by

$$\langle f, g \rangle \equiv \langle f | g \rangle = \int_{\mathbf{F}^n} \overline{f(x)} g(x) \mathrm{d}x.$$

Clearly, $P$ is linear and $T$ is anti-linear (conjugate linear) satisfying

$$P^2 = T^2 = I, PT = TP. \tag{2}$$

A Hamiltonian $H$ is said to be $PT$ **-symmetric** if it commutes with $PT$, i.e.,

$$[H, PT] = HPT - PTH = 0. \tag{3}$$

Since $(PT)^2 = I$, we see that $PT$-symmetry of $H$ is equivalent to

$$H^{PT} := (PT)H(PT) = H. \tag{4}$$

Every Hermitian Hamiltonian has real spectrum, but not every $PT$-symmetric Hamiltonian

has real spectrum. It was proved in [9] that if $H$ has unbroken $PT$-symmetry, then the eigenvalues of $H$ are all real. The key point of *PT*-symmetric quantum theory is to build a new positive definite inner product on the given Hilbert space so that the given Hamiltonian is Hermitian with respect to the new inner product. Hence, the Hilbert space in $PT$-symmetric quantum mechanics is formulated as a linear vector space with a dynamic inner product.

The aim of this note is to give some further mathematical discussions on this theory. Especially, concepts of *PT*-frames, *CPT*-frames on a Hilbert space and for a Hamiltonian will be proposed and the existence and construction of *CPT*-frames should be discussed.

## 2. *PT*-Frames and *PT*-symmetry

In this section, based on the Bender's idea, we will introduce and discuss the **abstract** $PT$-frames on a complex Hilbert space $K$ and the $PT$-symmetry of an operator $H$ in a Hilbert space. In what follows, we use $A^+$ to denote the Dirac adjoint of an operator $A$. Thus, $A$ is Hermitian if and only if $A = A^+$.

**Definition 2.1.** Let $P, T$ be continuous operators on a complex Hilbert space $K$ such that

(1) $P$ is linear and not the identity operator $I$ and $T$ is conjugate linear;

(2) $P^2 = T^2 = I$, $PT = TP$.

Then the pair $\{P, T\}$ is called a $PT$-**frame** on $K$. An operator $H : D(H) \to K$ is said to be $PT$-**symmetric** if it commutes with the operator $PT$. In this case, we also see that $H$ has $PT$-symmetry.

Note that $H$ commutes with the operator $PT$ means that
$$PT(D(H)) \subset D(H) \text{ and } PTHx = HPTx, \forall x \in D(H). \tag{5}$$

For a $PT$-frame on $K$ and an operator $H : K \supset D(H) \to K$, define $H^{PT} = PTHPT$. Then $H$ is $PT$-symmetric if and only if $H^{PT} = H$. Clearly, for every polynomial $p(x)$ with real coefficients, the operator $p(PT)$ is $PT$-symmetric. For a linear subspace $Y$ of $K$, we use $PT(Y, K)$ to denote the set of all $PT$-symmetric operators from $Y$ into $K$. Clearly, $PT(Y, K)$ is a real linear space, and $PT(K) := PT(K, K)$ is a real unital algebra.

Let $K$ be a complex Hilbert space with an orthonormal basis $\{e_i\}_{i=1}^{\infty}$. Define
$$P\sum_{i=1}^{\infty} x_i e_i = x_2 e_1 + x_1 e_2 + x_4 e_3 + x_3 e_4 + x_6 e_5 + x_5 e_6 + \cdots, \quad T\sum_{i=1}^{\infty} x_i e_i = \sum_{i=1}^{\infty} \overline{x_i} e_i,$$
then $\{P, T\}$ is a $PT$-frame on $K$. This shows that every Hilbert space has always a $PT$-frame. Moreover, define operators $P, T : K \oplus K \to K \oplus K$ as follows:
$$P(x, y) = (y, x), T(x, y) = (\overline{x}, \overline{y}), \quad \forall (x, y) \in K \oplus K,$$
where $\overline{x} = \sum_{i=1}^{\infty} \overline{\xi_i} e_i, \forall x = \sum_{i=1}^{\infty} \xi_i e_i \in K$. Clearly,
$$P = \begin{bmatrix} 0 & I_K \\ I_K & 0 \end{bmatrix}, \quad T = \begin{bmatrix} T_c & 0 \\ 0 & T_c \end{bmatrix}, \text{ where } T_c x = \overline{x}.$$

Then $\{P, T\}$ is a $PT$-frame on $K \oplus K$. For any densely defined linear operator $H : K \supset D(H) \to K$, put
$$\tilde{H} = \begin{bmatrix} H & 0 \\ 0 & H^+ \end{bmatrix} \equiv H \oplus H^+.$$

Then $PT\tilde{H}PT = TP\tilde{H}PT = T\tilde{H}^+T$, and so $\tilde{H}$ is $PT$-symmetric if and only if $T\tilde{H}^+T = \tilde{H}$ if and only if $T_c H^+ T_c = H$. A similar argument is also valid for finite dimensional case. Thus, for every complex matrix $H = [h_{ij}] \in M_n$, the operator $\tilde{H}$ is $PT$-symmetric on $\mathbb{C}^n \oplus \mathbb{C}^n$ if and only if $H$ is symmetric: $H^t := [h_{ji}] = [h_{ij}]$, where

$$P = \begin{bmatrix} 0 & I_n \\ I_n & 0 \end{bmatrix}, \quad T = \begin{bmatrix} T_c & 0 \\ 0 & T_c \end{bmatrix},$$

with $I_n$ is the $n \times n$ unit matrix and $T_c(x_1, x_2, \cdots, x_n) = (\overline{x_1}, \overline{x_2}, \cdots, \overline{x_n})$.

Next result gives the spectral properties of a $PT$-symmetric operator.

Recall that the resolvent $\rho(H)$ and the spectrum $\sigma(H)$ of a densely defined linear operator $H : K \supset D(H) \to K$ are defined by

$$\rho(H) = \{\lambda \in \mathbf{F} : (\lambda I_K - H)^{-1} \in B(K)\}, \quad \sigma(H) = \mathbf{F} \setminus \rho(H).$$

**Proposition 2.1**. *Let* $\{P, T\}$ *be a PT-frame on* $K$ *and* $H : K \supset D(H) \to K$ *be densely defined. If* $H$ *has PT-symmetry, then*

*(1)* $\lambda \in \sigma(H) \Leftrightarrow \overline{\lambda} \in \sigma(H)$;

*(2)* $\lambda \in \sigma_p(H) \Leftrightarrow \overline{\lambda} \in \sigma_p(H)$.

**Proof.** Recall that the resolvent and the spectrum of $H : K \supset D(H) \to K$ are defined as follows:

$$\rho(H) = \{\lambda \in \mathbf{C} : \overline{R(\lambda I - H)} = K, (\lambda I - H)^{-1} \in B(R(\lambda I - H), K)\}, \quad \sigma(H) = \mathbf{C} \setminus \rho(H),$$

where $B(R(\lambda I - H), K)$ is the set of all bounded linear operators from $R(\lambda I - H)$ into $K$. While the point spectrum $\sigma_p(H)$ of $H$ is defined as

$$\sigma_p(H) = \{\lambda \in \mathbf{C} : \ker(\lambda I - H) \neq \{0\}\},$$

the set of all eigenvalues of $H$.

Let $\lambda \in \rho(H)$. Then $\overline{R(\lambda I - H)} = K, (\lambda I - H)^{-1} \in B(R(\lambda I - H), K)$. Since $\overline{\lambda} I - H = PT(\lambda I - H)PT$ and $PT$ is a conjugate linear homeomorphism from $K$ onto $K$, we see that $\overline{R(\overline{\lambda} I - H)} = \overline{R(\lambda I - H)} = K$ and

$$(\overline{\lambda} I - H)^{-1} = PT(\lambda I - H)^{-1} PT \in B(R(\overline{\lambda} I - H), K).$$

This shows that $\overline{\lambda} \in \rho(H)$. So, $\lambda \in \rho(H) \Leftrightarrow \overline{\lambda} \in \rho(H)$, and then $\lambda \in \sigma(H) \Leftrightarrow \overline{\lambda} \in \sigma(H)$. Also, it follows from $\overline{\lambda} I - H = PT(\lambda I - H)PT$ that $\ker(\overline{\lambda} I - H) = \ker(\lambda I - H)$. Hence, $\lambda \in \sigma_p(H) \Leftrightarrow \overline{\lambda} \in \sigma_p(H)$. The proof is completed.

From Proposition 2.1, we know that the spectrum of $PT$-symmetric operator is necessarily real. See proposition 2.3 below.

**Definition 2.2.** Let $\{P, T\}$ be a $PT$-frame on $K$ and $H : K \supset D(H) \to K$. If $H$ is $PT$-symmetric and every eigenstate of $H$ is also an eigenstate of $PT$, then we say that the $PT$-symmetry of $H$ is unbroken, or $H$ has unbroken $PT$-symmetry.

**Proposition 2.2**. *If* $H$ *has unbroken PT-symmetry, then the eigenvalues of* $H$ *are all real.*

**Proof.** Let $a$ be any eigenvalue of $H$ with eigenstate $f \neq 0$. Then $Hf = af$. Thus, the $PT$-symmetry of $H$ implies that

$$HPTf = PTHf = \overline{a}PTf. \tag{10}$$

Since the vector $f$ is also an eigenstate of $PT$, there exists a complex number $b$ such that $PTf = bf$. By (10), we have $abf = HPTf = \overline{a}PTf = \overline{a}bf$. Since $b \neq 0$ and $f \neq 0$, $a = \overline{a}$. This shows that $a$ is real. The proof is completed.

Generally, we see from the proof above that when $H$ has unbroken $PT$-symmetry, if an eigenvalue $a$ of $H$ has an eigenstate which is also an eigenstate of $PT$, then $a$ must be real.

**Proposition 2.3.** *Let*

$$H = \begin{pmatrix} a & b \\ c & d \end{pmatrix} (a,b,c,d \in \mathbf{C}), \quad P = \begin{pmatrix} 0 & 1 \\ 1 & 0 \end{pmatrix}, \quad T\begin{pmatrix} x \\ y \end{pmatrix} = \begin{pmatrix} \bar{x} \\ \bar{y} \end{pmatrix}.$$

Then $\{P,T\}$ is a *PT*-frame on $\mathbf{C}^2$, and the following conclusions are valid.

(1) $THT = \bar{H} := \begin{pmatrix} \bar{a} & \bar{b} \\ \bar{c} & \bar{d} \end{pmatrix}$ (complex conjugate).

(2) $H^+ := (\bar{H})^t = (THT)^t$ ($t$ denotes the transpose of a matrix); $H^T := TH^+T = H^t$.

(3) $H^+ = H \Leftrightarrow TH = H^T T \Leftrightarrow TH = H^+ T \Leftrightarrow H = \begin{pmatrix} a & b \\ \bar{b} & d \end{pmatrix} (a,d \in \mathbf{R})$.

(4) $H^T = H \Leftrightarrow H^t = H \Leftrightarrow H = \begin{pmatrix} a & b \\ b & d \end{pmatrix}$.

(5) $H^{PT} = H \Leftrightarrow PH = \bar{H}P \Leftrightarrow H = \begin{pmatrix} a & b \\ \bar{b} & \bar{a} \end{pmatrix}$.

(6) $H^{PT} = H = H^+ \Leftrightarrow H = \begin{pmatrix} a & b \\ \bar{b} & a \end{pmatrix} (a \in \mathbf{R})$.

(7) $H^{PT} = H = H^+ = H^T \Leftrightarrow H = \begin{pmatrix} a & b \\ b & a \end{pmatrix} (a,b \in \mathbf{R})$.

Consider the following matrices:

$$H_1 = \begin{pmatrix} 1 & 2 \\ 2 & 3 \end{pmatrix}, \quad H_2 = \begin{pmatrix} 1+i & 2i \\ -2i & 1-i \end{pmatrix}, \quad H_3 = \begin{pmatrix} 1 & 2i \\ -2i & 3 \end{pmatrix}.$$

From Definition 2.1 and Proposition 2.3, we know that with respect to the *PT*-frame $\{P,T\}$ given by Proposition 2.4:

(1) A real symmetric matrix is not necessarily *PT*-symmetric, e.g., $H_1$.
(2) A *PT*-symmetric matrix is not necessarily symmetric, e.g., $H_2$.
(3) A Hermitian matrix is not necessarily *PT*-symmetric, e.g., $H_3$.

**Example 2.1.** Let $\{P,T\}$ be the *PT*-frame given by Proposition 2.2 and

$$H = \begin{pmatrix} re^{i\theta} & s \\ s & re^{-i\theta} \end{pmatrix}, \tag{11}$$

where $r, s, \theta$ are nonzero real numbers. It is easy to check that $H$ is *PT*-symmetric and non-Hermitian.

When $|r/s \cdot \sin\theta| \leq 1$, $H$ has two real eigenvalues:

$$E_{\pm} = r\cos\theta \pm s\cos\varphi \, (\sin\varphi = r/s \cdot \sin\theta),$$

with the corresponding eigenstates $|\psi_{\pm}\rangle$, where

$$|\psi_+\rangle = \frac{1}{\sqrt{2}}\begin{pmatrix} e^{i\varphi/2} \\ e^{-i\varphi/2} \end{pmatrix}, |\psi_-\rangle = \frac{1}{\sqrt{2}}\begin{pmatrix} e^{-i\varphi/2} \\ -e^{i\varphi/2} \end{pmatrix}.$$

Clearly, $PT|\psi_{\pm}\rangle = \pm|\psi_{\pm}\rangle$ and so $H$ has unbroken *PT*-symmetry.

When $|r/s \cdot \sin\theta| > 1$, $H$ has two non-real eigenvalues:

$$E_{\pm} = r\cos\theta \pm i\sqrt{r^2\sin^2\theta - s^2}.$$

Thus, $H$ has no unbroken *PT*-symmetry.

**Example 2.2.** Let

$$H = \begin{pmatrix} re^{i\theta} & s & 0 \\ s & re^{-i\theta} & 0 \\ 0 & 0 & a \end{pmatrix}, P = \begin{pmatrix} 0 & 1 & 0 \\ 1 & 0 & 0 \\ 0 & 0 & 1 \end{pmatrix}, T\begin{pmatrix} x \\ y \\ z \end{pmatrix} = \begin{pmatrix} \bar{x} \\ \bar{y} \\ \bar{z} \end{pmatrix},$$

where $r, s, \theta$ and $a$ are nonzero real numbers. It is easy to check that $\{P, T\}$ becomes a $PT$-frame on $\mathbf{C}^3$ and $H$ is $PT$-symmetric and non-Hermitian. When $|r/s \cdot \sin\theta| \leq 1$, $H$ has three real eigenvalues:

$$a \text{ and } E_\pm = r\cos\theta \pm s\cos\varphi \, (\sin\varphi = r/s \cdot \sin\theta),$$

with the corresponding eigenstates $|\psi_a\rangle$ and $|\psi_\pm\rangle$, where

$$|\psi_a\rangle = \begin{pmatrix} 0 \\ 0 \\ 1 \end{pmatrix}, |\psi_+\rangle = \frac{1}{\sqrt{2}}\begin{pmatrix} e^{i\varphi/2} \\ e^{-i\varphi/2} \\ 0 \end{pmatrix}, |\psi_-\rangle = \frac{1}{\sqrt{2}}\begin{pmatrix} e^{-i\varphi/2} \\ -e^{i\varphi/2} \\ 0 \end{pmatrix}.$$

Clearly, $PT|\psi_a\rangle = |\psi_a\rangle$ and $PT|\psi_\pm\rangle = \pm|\psi_\pm\rangle$, and so $H$ has unbroken $PT$-symmetry.

When $|r/s \cdot \sin\theta| > 1$, $H$ has a real eigenvalue $a$ and two non-real eigenvalues:

$$E_\pm = r\cos\theta \pm i\sqrt{r^2\sin^2\theta - s^2}.$$

Thus, $H$ has no unbroken $PT$-symmetry.

**Example 2.3.** Let

$$H = \begin{pmatrix} r_1 e^{i\theta_1} & s_1 & 0 & 0 \\ s_1 & r_2 e^{-i\theta_2} & 0 & 0 \\ 0 & 0 & r_2 e^{i\theta_2} & s_2 \\ 0 & 0 & s_2 & r_2 e^{-i\theta_2} \end{pmatrix}, P = \begin{pmatrix} 0 & 1 & 0 & 0 \\ 1 & 0 & 0 & 0 \\ 0 & 0 & 0 & 1 \\ 0 & 0 & 1 & 0 \end{pmatrix}, T\begin{pmatrix} x \\ y \\ z \\ u \end{pmatrix} = \begin{pmatrix} \bar{x} \\ \bar{y} \\ \bar{z} \\ \bar{u} \end{pmatrix}, \quad (12)$$

Where $r_k, s_k, \theta_k$ are nonzero real numbers. It is easy to check that $\{P, T\}$ becomes a $PT$-frame on $\mathbf{C}^4$ and $H$ is $PT$-symmetric and non-Hermitian. It can be computed that when $|r_k/s_k \cdot \sin\theta_k| \leq 1 (k=1,2)$, $H$ has four real eigenvalues:

$$E_\pm^k = r_k\cos\theta_k \pm s_k\cos\varphi_k \, (\sin\varphi_k = r_k/s_k \cdot \sin\theta_k),$$

with the corresponding eigenstates $|\psi_+^1\rangle, |\psi_-^1\rangle, |\psi_+^2\rangle, |\psi_-^2\rangle$, where

$$|\psi_+^1\rangle = \frac{1}{\sqrt{2}}\begin{pmatrix} e^{i\varphi_1/2} \\ e^{-i\varphi_1/2} \\ 0 \\ 0 \end{pmatrix}, |\psi_-^1\rangle = \frac{1}{\sqrt{2}}\begin{pmatrix} e^{-i\varphi_1/2} \\ -e^{i\varphi_1/2} \\ 0 \\ 0 \end{pmatrix},$$

$$|\psi_+^2\rangle = \frac{1}{\sqrt{2}}\begin{pmatrix} 0 \\ 0 \\ e^{i\varphi_2/2} \\ e^{-i\varphi_2/2} \end{pmatrix}, |\psi_-^2\rangle = \frac{1}{\sqrt{2}}\begin{pmatrix} 0 \\ 0 \\ e^{-i\varphi_2/2} \\ -e^{i\varphi_2/2} \end{pmatrix}.$$

Since $PT|\psi_\pm^k\rangle = \pm|\psi_\pm^k\rangle \, (k=1,2)$, $H$ has unbroken $PT$-symmetry.

When $|r_1/s_1 \cdot \sin\theta_1| > 1$, or $|r_2/s_2 \cdot \sin\theta_2| > 1$, $H$ has two non-real eigenvalues:

$$E_\pm^1 = r_1\cos\theta_1 \pm i\sqrt{r_1^2\sin^2\theta_1 - s_1^2}, \text{ or}$$

$$E_\pm^2 = r_2\cos\theta_2 \pm i\sqrt{r_2^2\sin^2\theta_2 - s_2^2}.$$

Thus, $H$ has no unbroken $PT$-symmetry.

**Example 2.4.** Let

$$H_k = \begin{pmatrix} r_k e^{i\theta_k} & s_k \\ s_k & r_k e^{-i\theta_k} \end{pmatrix}, \quad P_k = \begin{pmatrix} 0 & 1 \\ 1 & 0 \end{pmatrix}, \quad T_k \begin{pmatrix} x \\ y \end{pmatrix} = \begin{pmatrix} \bar{x} \\ \bar{y} \end{pmatrix}, \quad (13)$$

where $r_k, s_k, \theta_k$ are nonzero real numbers ($k=1,2,\cdots$) such that the sequences $\{r_k\}$ and $\{s_k\}$ are bounded. Put

$$H = \bigoplus_{k=1}^{\infty} H_k, \quad P = \bigoplus_{k=1}^{\infty} P_k, \quad T = \bigoplus_{k=1}^{\infty} T_k.$$

Then $H, P$ are bounded linear operators on the Hilbert space:

$$\ell^2 = \ell^2(\mathbf{N}) = \{(x_1, x_2, \cdots, x_n, \cdots) : x_k \in \mathbf{C}(\forall k), \sum_{k=1}^{\infty} |x_k|^2 < \infty\}$$

while $T$ is a bounded conjugate linear operator on the same space satisfying

$$P^2 = T^2 = I, PT = TP, HPT = PTH.$$

Clearly, $\{P, T\}$ becomes a $PT$-frame on $\ell^2$ and $H$ is $PT$-symmetric.

From Example 2.1, we see that when $|r_k/s_k \cdot \sin\theta_k| \leq 1 (k=1,2,\cdots)$, $H$ has real eigenvalues: $E_+^1, E_-^1, E_+^2, E_-^2, \cdots, E_+^n, E_-^n, \cdots$, where $E_\pm^k = r_k \cos\theta_k \pm s_k \cos\varphi_k$ and $\sin\varphi_k = r_k/s_k \cdot \sin\theta_k$, with the corresponding eigenvalues: $\alpha_+^1, \alpha_-^1, \alpha_+^2, \alpha_-^2, \cdots, \alpha_+^n, \alpha_-^n, \cdots$, where

$$\alpha_+^1 = (\psi_+^1, 0, 0, 0, \cdots, 0, \cdots),$$
$$\alpha_-^1 = (\psi_-^1, 0, 0, 0, \cdots, 0, \cdots),$$
$$\alpha_+^2 = (0, 0, \psi_+^2, 0, 0, 0, \cdots, 0, \cdots),$$
$$\alpha_-^2 = (0, 0, \psi_-^2, 0, 0, 0, \cdots, 0, \cdots),$$
$$\cdots\cdots$$
$$\alpha_+^n = (\overbrace{0, 0, \cdots, 0, 0}^{2(n-1)}, \psi_+^n, 0, 0, 0, \cdots, 0, \cdots),$$
$$\alpha_-^n = (\overbrace{0, 0, \cdots, 0, 0}^{2(n-1)}, \psi_-^n, 0, 0, 0, \cdots, 0, \cdots),$$
$$\cdots\cdots,$$

where

$$\psi_+^k = \frac{1}{\sqrt{2}}\left(e^{i\varphi_k/2}, e^{-i\varphi_k/2}\right), \psi_-^k = \frac{1}{\sqrt{2}}\left(e^{-i\varphi_k/2}, -e^{i\varphi_k/2}\right).$$

Clearly, $\alpha_+^1, \alpha_-^1, \alpha_+^2, \alpha_-^2, \cdots, \alpha_+^n, \alpha_-^n, \cdots$ are eigenstates of $PT$ for eigenvalues

$$1, -1, 1, -1, \cdots, 1, -1, \cdots.$$

Hence, $H$ has unbroken $PT$-symmetry. In the case where $|r_k/s_k \cdot \sin\theta_k| > 1$ for some $k$, $H$ has two non-real eigenvalues:

$$E_\pm^k = r_k \cos\theta_k \pm i\sqrt{r_k^2 \sin^2\theta_k - s_k^2}.$$

Therefore, $H$ has no unbroken $PT$-symmetry.

**Example 2.5.** For every $f \in L^2(\mathbf{R}^2)$, define

$$(Pf)(x,y) = f(-x,-y), \quad (Tf)(x,y) = \overline{f(x,y)},$$
$$(P_1 f)(x,y) = f(-x,y), \quad (P_2 f)(x,y) = f(x,-y).$$

Then $P^2 = I = T^2, P_k^2 = I, P_k T = TP_k (k=1,2), PT = TP$ and so the pairs $\{P, T\}$ and

$\{P_k, T\}(k=1,2)$ are all $PT$-frames on $L^2(\mathbf{R}^2)$. Clearly, $P := P_1 P_2 = P_2 P_1$. It is easy to see that the Hamiltonian

$$H = \frac{1}{2}(\hat{p}^2 + \hat{q}^2) + \frac{1}{2}(\hat{x}^2 + \hat{y}^2) + i\varepsilon \hat{x}^3 \hat{y}^3 (\varepsilon \in \mathbf{R})$$

is both $P_1 T$-symmetric and $P_2 T$-symmetric, but not $PT$-symmetric.

**Example 2.6.** For every $f \in L^2(\mathbf{R}^3)$, define

$$(Pf)(x,y,z) = f(-x,-y,-z), \quad (P_1 f)(x,y,z) = f(-x,y,z),$$
$$(P_2 f)(x,y,z) = f(x,-y,z), \quad (P_3 f)(x,y,z) = f(x,y,-z),$$
$$(P_4 f)(x,y,z) = f(-x,-y,z), \quad (Tf)(x,y,z) = \overline{f(x,y,z)}.$$

Then the pairs $\{P, T\}$ and $\{P_k, T\}(k=1,2,3,4)$ are all $PT$-frames on $L^2(\mathbf{R}^3)$. Clearly, $P = P_1 P_2 P_3 = P_3 P_2 P_1 = P_2 P_1 P_3$. It is easy to see that the Hamiltonian

$$H = \frac{1}{2}(\hat{p}^2 + \hat{q}^2 + \hat{r}^2) + \frac{1}{2}(\hat{x}^2 + \hat{y}^2 + \hat{z}^2) + i\varepsilon \hat{x}^3 \hat{y}^3 \hat{z}^3 (\varepsilon \in \mathbf{R})$$

is both $PT$-symmetric and $P_k T(k=1,2,3)$-symmetric, but not $P_4 T$-symmetric since $H \cdot P_4 T \neq P_4 T \cdot H$.

## 3. *CPT*-Frames

In this section, based on the Bender's idea [9], we will introduce and discuss the abstract *CPT*-frames on a complex Hilbert space $K$ and for an operator $H$ in a Hilbert space.

**Definition 3.1.** Let $\{P, T\}$ be a $PT$-frame on a Hilbert space $K$, and $C$ a linear operator on $K$ such that

(1) $CPT = TPC$, $C^2 = I$;

(2) $PC$ is positive definite with respective to the inner product $\langle \cdot, \cdot \rangle$ on $K$, i.e.,

$$\langle PCx, x \rangle \geq 0 (\forall x \in K); \quad \langle PCx, x \rangle = 0 \Rightarrow x = 0.$$

Then the triple $\{C, P, T\}$ is said to be a *CPT*-frame on $K$. If in addition, $C$ commutes with $H : K \supset D(H) \to K$, then the triple $\{C, P, T\}$ is said to be a *CPT*-frame for $H$.

**Proposition 3.1.** *Let* $\{C, P, T\}$ *be a CPT-frame on* $K$.

*(1) The formula* $\langle \psi, \varphi \rangle_{CPT} = \langle PC\psi, \varphi \rangle_K$ *defines a positive definite inner product on* $K$, *called a CPT-inner product on* $K$. *Moreo0ver, the norm* $\|\cdot\|_{CPT}$ *induced by this new inner product satisfies*

$$\|CP\|^{-1/2} \|\psi\| \leq \|\psi\|_{CPT} \leq \|PC\|^{1/2} \|\psi\|, \quad \forall \psi \in K.$$

*(2) With respect to CPT-product, the adjoint operator of a densely defined linear operator* $A : K \supset D(A) \to K$ *is* $A^{CPT} = (PC)^{-1} A^+ (PC)$, *where* $A^+$ *denotes the usual adjoint of* $A$ *with respect to the original inner product* $\langle \cdot, \cdot \rangle$ *on* $K$.

*(3)* $A^{CPT} = (CPT)^{-1} A^T (CPT)$, *where* $A^T := TA^+ T$ *(called the transpose of* $A$*)*.

*(4)* $A^{CPT} = A \Leftrightarrow PCA = A^+ PC$; *when* $A^+ = A$, $A^{CPT} = A \Leftrightarrow [A, PC] = 0$.

*(5) When* $A^T = A$, $A^{CPT} = A \Leftrightarrow [A, CPT] = 0$.

*(6) If* $\{C, P, T\}$ *is a CPT-frame for a PT-symmetric operator* $H : K \to K$ *and* $h = (PC)^{1/2} H(PC)^{-1/2}$, *then* $h^+ = h$ *if and only if* $H$ *is symmetric ( i.e.,* $H^T = H$ *) if and only if* $H$ *is CPT-Hermitian, i.e.,* $H^{CPT} = H$.

**Proof.** (1) Clearly. $\forall \psi \in D(A^{CPT}), \varphi \in D(A)$, we compute that

$$\langle A^{CPT}\psi,\varphi\rangle_{CPT} = \langle \psi, A\varphi\rangle_{CPT} = \langle PC\psi, A\varphi\rangle_K$$
$$= \langle A^+PC\psi,\varphi\rangle_K = \langle (PC)^{-1}A^+PC\psi,\varphi\rangle_{CPT}.$$

So, $A^{CPT} = (PC)^{-1}A^+(PC)$ and (2) follows. Since $A^T = TA^+T$ and $T^2 = I$, we see $A^+ = TA^TT$ and then (2) implies that

$$A^{CPT} = (PC)^{-1}TA^TT(PC) = (T \cdot PC)^{-1}A^T(C \cdot TP) = (CPT)^{-1}A^T(CPT).$$

Conclusions (4) and (5) can be obtained by (2) and (3), respectively. To see (6), we note that $H^{PT} = H$ implies that $THT = PHP$. Thus, $H^T = H$ if and only if $H^+ = PHP$. Since $(PC)^+ = PC$, $P^{-1} = P, C^{-1} = C, HC = CH$, we get

$$h^+ = (PC)^{-1/2}H^+(PC)^{1/2},$$
$$h = (PC)^{-1/2}(PC)H(PC)^{-1}(PC)^{1/2} = (PC)^{-1/2}(PHP)(PC)^{1/2}.$$

Hence, $h^+ = h$ if and only if $H^+ = PHP$ if and only if $H^T = H$. By (1), we see that

$$H^{CPT} = H \Leftrightarrow (CPT)H = H^T(CPT) \Leftrightarrow H(CPT) = H^T(CPT) \Leftrightarrow H = H^T.$$

The proof is completed.

**Corollary 3.1.** *Let operators* $P,T$ *be as in Eqn. (1) and* $\{C,P,T\}$ *a CPT-frame for a PT-symmetric operator* $H : D(H) \to L^2(\mathbf{F}^n)$. *Then*

$$\langle \psi,\varphi\rangle_{CPT} := \langle \psi|PC|\varphi\rangle_{L^2} = \int_{\mathbf{F}^n}\psi^{CPT}(x)\varphi(x)\mathrm{d}x \quad (\psi^{CPT} := CPT\psi), \qquad (14)$$

*$H$ is CPT-Hermitian if and only if $H$ is symmetric ($H^T := TH^+T = H$), and in that case, $H$ is similar to the operator $h = (PC)^{1/2}H(PC)^{-1/2}$ with $h^+ = h$, while different eigenvalues $E_n$ of $H$ have CPT-orthogonal eigenstates $\psi_n$.*

**Corollary 3.2.** *Let* $\{C,P,T\}$ *be a CPT on* $\mathbf{C}^n$, *where*

$$T|\psi\rangle = \langle\psi|^{tr} = |\psi\rangle^* \text{ (complex conjugate)}, \quad \forall |\psi\rangle, |\varphi\rangle \in \mathbf{C}^n.$$

*Define* $\langle\psi|_{CPT} = \langle\psi|PC$ *and* $\langle\psi,\varphi\rangle_{CPT} = \langle\psi,PC\varphi\rangle_{\ell^2} = \langle\psi|_{CPT}|\varphi\rangle$, *then*

(1) $\langle\cdot,\cdot\rangle_{CPT}$ *is a positive definite inner product on* $\mathbf{C}^n$.

(2) *The adjoint of a matrix (as an operator)* $A : \mathbf{C}^n \to \mathbf{C}^n$ *with respect to* $\langle\cdot,\cdot\rangle_{CPT}$ *is*

$$A^{CPT} = (PC)^{-1}A^+(PC), \text{ where } A^+ = (\overline{A})^T.$$

(3) $A^T := TA^+T = A^{tr}$, $A^{CPT} = (CPT)^{-1}A^{tr}(CPT)$.

(4) $A^{CPT} = A \Leftrightarrow PCA = A^+PC$; *when* $A^+ = A$, $A^{CPT} = A \Leftrightarrow [A,PC] = 0$.

(5) *When* $A^T = A$, $A^{CPT} = A \Leftrightarrow [A,CPT] = 0$.

(6) *If* $H \in M_n(\mathbf{C})$ *is symmetric* ($H^T = H$) *and PT-symmetric such that* $HC = CH$, *then $H$ is CPT-Hermitian and similar to Hermitian matrix* $h = (PC)^{1/2}H(PC)^{-1/2}$.

To discuss the existence of $CPT$-frames on $L^2(\mathbf{F}^n)$, we need the following lemma.

**Lemma 3.1.** *Let* $f_n$ *be Lebesgue measurable functions on* $E \subset \mathbf{R}^n$ *($n = 0,1,2,\cdots$) with* $\int_E \sum_{n=0}^{\infty} |f_n(x)|\mathrm{d}x < \infty$. *Then the series* $\sum_{n=0}^{\infty} f_n(x)$ *is almost everywhere convergent on $E$ and its sum function is Lebesgue integrable on $E$ with*

$$\int_E \sum_{n=0}^{\infty} f_n(x)\mathrm{d}x = \sum_{n=0}^{\infty}\int_E f_n(x)\mathrm{d}x.$$

**Theorem 3.2.** *Let* $P,T$ *be as in Eqn. (1) and $C$ a bounded linear operator on* $L^2(\mathbf{F}^n)$

*defined by*

$$(C\psi)(x) = \int_{\mathbf{F}^n} c(x,y)\psi(y)\mathrm{d}y, \qquad (15)$$

*where $c(x,y)$ is an integral kernel. Then*

*(1) $PTC = CPT$ if and only if $(c(x,y))^* = c(-x,-y)$, $\forall a.e.\ x, y \in \mathbf{F}^n$.*

*(2) $C^2 = I$ if and only if $\psi(z) = \int_{\mathbf{F}^n} f(z,y)\psi(y)\mathrm{d}y, \forall \psi \in L^2(\mathbf{F}^n)$ and $\forall a.e.\ z \in \mathbf{F}^n$ where*

$$f(z,y) = \int_{\mathbf{F}^n} c(x,y)c(z,x)\mathrm{d}x.$$

*(3) If $\{\varphi_n\}_{n\in\mathbf{N}}$ is an orthonormal basis for $L^2(\mathbf{F}^n)$ with $\varphi_n(-x) = \varphi_n(x) = \overline{\varphi_n(x)}$ for all $n$ and all $x \in E$ such that $\sum_{n=0}^{\infty}\|\varphi_n\|_1 < \infty$, and $c(x,y) = \sum_{n=0}^{\infty}\varphi_n(x)\varphi_n(y)$, then $\{C, P, T\}$ is a CPT-frame on $L^2(\mathbf{F}^n)$.*

**Proof.** (1) By Eqn. (15), we compute that $\forall \psi \in L^2(\mathbf{F}^n)$,

$$(PTC\psi)(x) = \int_{\mathbf{F}^n} (c(-x,y))^* (\psi(y))^* \mathrm{d}y = \int_{\mathbf{F}^n} (c(-x,-y))^* (\psi(-y))^* \mathrm{d}y,$$

and

$$(CPT\psi)(x) = \int_{\mathbf{F}^n} c(x,y)(\psi(-y))^* \mathrm{d}y.$$

Thus, $PTC = CPT$ if and only if

$$\int_{\mathbf{F}^n} (c(-x,-y))^* (\psi(-y))^* \mathrm{d}y = \int_{\mathbf{F}^n} c(x,y)(\psi(-y))^* \mathrm{d}y \quad (\forall \psi \in L^2(\mathbf{F}^n), a.e.\ x \in \mathbf{F}^n)$$

if and only if $(c(x,y))^* = c(-x,-y), \forall a.e.\ x, y \in \mathbf{F}^n$.

(2) $\forall \psi \in L^2(\mathbf{F}^n)$, we have $(C\psi)(x) = \int_{\mathbf{F}^n} c(x,y)\psi(y)\mathrm{d}y$ and so

$$(C^2\psi)(z) = \int_{\mathbf{F}^n} \mathrm{d}x \int_{\mathbf{F}^n} c(z,x)c(x,y)\psi(y)\mathrm{d}y$$

$$= \int_{\mathbf{F}^n} \psi(y)\mathrm{d}y \int_{\mathbf{F}^n} c(z,x)c(x,y)\mathrm{d}x$$

$$= \int_{\mathbf{F}^n} \psi(y)f(z,y)\mathrm{d}y,$$

where $f(z,y) = \int_{\mathbf{F}^n} c(z,x)c(x,y)\mathrm{d}x.$ Hence,

$$C^2 = I \text{ if and only if } (C^2\psi)(z) = \psi(z), \ \forall \psi \in L^2(\mathbf{F}^n), \forall a.e.\ z \in \mathbf{F}^n.$$

The later is equivalent to

$$\psi(z) = \int_{\mathbf{F}^n} f(z,y)\psi(y)\mathrm{d}y, \forall \psi \in L^2(\mathbf{F}^n), \ \forall a.e.\ z \in \mathbf{F}^n.$$

(3) $\forall \psi \in L^2(\mathbf{F}^n)$, the condition that $\sum_{n=0}^{\infty}\|\varphi_n\|_1 < \infty$ yields that

$$\int_{\mathbf{F}^n} \sum_{n=0}^{\infty} \int_{\mathbf{F}^n} |\varphi_n(x)\varphi_n(y)\psi(y)|\mathrm{d}y\mathrm{d}x \leq \sum_{n=0}^{\infty} \|\varphi_n\|_1 \cdot \|\psi\|_2 < \infty.$$

Thus,

$$\sum_{n=0}^{\infty} \int_{\mathbf{F}^n} |\varphi_n(x)\varphi_n(y)\psi(y)|\mathrm{d}y < \infty, \quad a.e.\ x \in \mathbf{F}^n.$$

Lemma 3.1 implies that, $\forall a.e.\ x \in \mathbf{F}^n$,

$$(C\psi)(x) = \int_{\mathbf{F}^n} c(x,y)\psi(y)\mathrm{d}y = \int_{\mathbf{F}^n} \sum_{n=0}^{\infty} \varphi_n(x)\varphi_n(y)\psi(y)\mathrm{d}y = \sum_{n=0}^{\infty} \langle \varphi_n, \psi \rangle \varphi_n(x) \quad (16)$$

is valid. Since $\forall \psi \in L^2(\mathbf{F}^n)$, we have

$$\begin{aligned}
\langle PC\psi, \psi \rangle_{L^2} &= \int_{\mathbf{F}^n} (TPC\psi)(x)\psi(x)\mathrm{d}x \\
&= \int_{\mathbf{F}^n} (CTP\psi)(x)\psi(x)\mathrm{d}x \\
&= \int_{\mathbf{F}^n}\int_{\mathbf{F}^n} c(x,y)(\psi(-y))^*\psi(x)\mathrm{d}x\mathrm{d}y \\
&= \sum_{n=0}^{\infty} \int_{\mathbf{F}^n}\int_{\mathbf{F}^n} \varphi_n(x)\varphi_n(y)(\psi(-y))^*\psi(x)\mathrm{d}x\mathrm{d}y \\
&= \sum_{n=0}^{\infty} \int_{\mathbf{F}^n} \varphi_n(x)\psi(x)\mathrm{d}x \int_{\mathbf{F}^n} \varphi_n(y)(\psi(-y))^*\mathrm{d}y \\
&= \sum_{n=0}^{\infty} \int_{\mathbf{F}^n} \varphi_n(x)\psi(x)\mathrm{d}x \int_{\mathbf{F}^n} \varphi_n(-y)(\psi(y))^*\mathrm{d}y \\
&= \sum_{n=0}^{\infty} \int_{\mathbf{F}^n} \varphi_n(x)\psi(x)\mathrm{d}x \int_{\mathbf{F}^n} [\varphi_n(y)\psi(y)]^*\mathrm{d}y \\
&= \sum_{n=0}^{\infty} \left| \int_{\mathbf{F}^n} \varphi_n(x)\psi(x)\mathrm{d}x \right|^2 \\
&\geq 0,
\end{aligned}$$

we see that $PC$ is positive. On the other hand,

$$\langle PC\psi, \psi \rangle_{L^2} = \sum_{n=0}^{\infty} \left| \int_{\mathbf{F}^n} \varphi_n(x)\psi(x)\mathrm{d}x \right|^2 = \sum_{n=0}^{\infty} \left| \langle \varphi_n, \psi \rangle_{L^2} \right|^2,$$

this implies that $\langle PC\psi, \psi \rangle_{L^2} = 0$ if and only if $\langle \varphi_n, \psi \rangle_{L^2} = 0 (\forall n)$ if and only if $\psi = 0$ since $\overline{\mathrm{span}\{\varphi_n\}_{n \in N}} = L^2(\mathbf{F}^n)$. Now, we have proved that $PC$ positive definite. By using the assumption that $\varphi_n(-x) = \varphi_n(x) = (\varphi_n(x))^*$, we get that

$$TP \cdot C\varphi_m = PT \cdot C\varphi_m = C \cdot PT\varphi_m, \forall m.$$

and so $TP \cdot C\varphi = PT \cdot C\varphi = C \cdot PT\varphi, \forall \varphi \in L^2(\mathbf{F}^n)$. That is, $[C, PT] = 0$. Since $\{\varphi_n\}_{n \in \mathbf{N}}$ is an orthonormal basis for $L^2(\mathbf{F}^n)$, Eqn. (16) implies that $C\psi = \sum_{n=0}^{\infty} \langle \varphi_n, \psi \rangle \varphi_n = \psi$ for all $\psi \in L^2(\mathbf{F}^n)$. So, $C = I$. The proof is completed.

Next theorem is about the existence of a $CPT$-frame for an unbroken $PT$-symmetric Hamiltonian.

**Theorem 3.3.** *Let $P, T$ be as in Eqn. (1), $M \leq L^2(\mathbf{R})$, and $H$ be a linear operator on $M$ and have unbroken $PT$-symmetry, whose normalized eigenstates $\{\psi_n\}$ generates $M$. Then there exist $\theta_n, E_n \in \mathbf{R}$ such that $\varphi_n := e^{i\theta_n/2}\psi_n$ satisfy $PT\varphi_n = \varphi_n$ and $H\varphi_n = E_n\varphi_n$ for all $n$. If in addition,*

$$(\varphi_m, \varphi_n) := \int_{\mathbf{R}} \varphi_m(x)\varphi_n(x)\mathrm{d}x = (-1)^n \delta_{m,n}, \; \sum_{n=0}^{\infty} \|\varphi_n\|_1 < \infty, \quad (17)$$

*then*

*(1) The function $c(x,y) = \sum_{n=0}^{\infty} \varphi_n(x)\varphi_n(y)$ is a.e. defined on $\mathbf{R}^2$ and measurable.*

*(2) The formula*

$$(C\psi)(x) = \int_{\mathbf{R}} c(x,y)\psi(y)\mathrm{d}y \quad (18)$$

*defines a linear operator $C$ on $M$.*

*(3) The triple $\{C,P,T\}$ is a $CPT$-frame for the operator $H$ and the eigenstates $\{\varphi_n\}_{n\in\mathbf{N}}$ are $CPT$-orthonormal.*

**Proof.** Since $\psi_n$'s are eigenstates of $H$, there are constants such that $H\psi_n = E_n\psi_n$. The unbroken $PT$-symmetry of $H$ implies that there are constants $c_n$ such that $PT\psi_n = c_n\psi_n$. The fact that $(PT)^2 = I$ implies that $|c_n|=1$. Thus, there exists $\theta_n \in \mathbf{R}$ such that $c_n = e^{i\theta_n}$. Clearly, $\varphi_n := e^{i\theta_n/2}\psi_n$ satisfies $PT\varphi_n = \varphi_n$ and $H\varphi_n = E_n\varphi_n$. Proposition 2.1 implies that $E_n$ are all real.

(1) The condition $\sum_{n=0}^{\infty}\|\varphi_n\|_1 < \infty$ ensures $\sum_{n=0}^{\infty}\|\varphi_n\|_1^2 < \infty$, and then

$$\int_{\mathbf{R}^2} \sum_{n=0}^{\infty} |\varphi_n(x)\varphi_n(y)| dxdy = \sum_{n=0}^{\infty}\left(\int_{\mathbf{R}}|\varphi_n(x)|dx\right)^2 = \sum_{n=0}^{\infty}\|\varphi_n\|_1^2 < \infty.$$

This shows that series $\sum_{n=0}^{\infty}\varphi_n(x)\varphi_n(y)$ is convergent almost everywhere on $\mathbf{R}^2$ and therefore its sum function $c(x,y)$ is a.e. defined and measurable on $\mathbf{R}^2$.

(2) Let $\psi \in M$. Then by $\sum_{n=0}^{\infty}\|\varphi_n\|_1 < \infty$, we have

$$\int_{\mathbf{R}} dx \int_{\mathbf{R}} |c(x,y)\psi(y)| dy \le \sum_{n=0}^{\infty} \int_{\mathbf{R}} |\varphi_n(x)| dx \int_{\mathbf{R}} |\varphi_n(y)\psi(y)| dy \le \sum_{n=0}^{\infty}\|\varphi_n\|_1\|\psi\|_2 < \infty.$$

Thus, $\int_{\mathbf{R}} |c(x,y)\psi(y)|dy < \infty, \quad \forall a.e.\ x \in \mathbf{R}$. This shows that the function $C\psi$ in (18) is well-defined on $\mathbf{R}$. Since

$$\int_{\mathbf{R}} \sum_{n=0}^{\infty} \int_{\mathbf{R}} |\varphi_n(x)\varphi_n(y)\psi(y)| dydx \le \sum_{n=0}^{\infty}\|\varphi_n\|_1\|\psi\|_2 < \infty,$$

we get that $\sum_{n=0}^{\infty} \int_{\mathbf{R}} |\varphi_n(x)\varphi_n(y)\psi(y)| dy < \infty, \quad a.e.\ x \in \mathbf{R}$. Therefore, Lemma 3.1 yields that

$$(C\psi)(x) = \int_{\mathbf{R}} c(x,y)\psi(y)dy = \int_{\mathbf{R}} \sum_{n=0}^{\infty}\varphi_n(x)\varphi_n(y)\psi(y)dy = \sum_{n=0}^{\infty}\int_{\mathbf{R}}\varphi_n(x)\varphi_n(y)\psi(y)dy.$$

Especially, when $\psi = \varphi_k$, (17) ensures that

$$(C\varphi_k)(x) = \sum_{n=0}^{\infty}\varphi_n(x)\int_{\mathbf{R}}\varphi_n(y)\varphi_k(y)dy = (-1)^k\varphi_k(x),\quad a.e.\ x \in \mathbf{R},$$

Thus, $C\varphi_k = (-1)^k\varphi_k\ (k=0,1,2,\cdots)$. This shows that the formula (18) defines an linear operator $C$ on $M$.

(3) $\forall m$, we compute from (2) that
$$C^2\varphi_m = C(-1)^m\varphi_m = (-1)^m C\varphi_m = (-1)^{2m}\varphi_m = \varphi_m,$$
$$HC\varphi_m = H(-1)^m\varphi_m = (-1)^m E_m\varphi_m = CH\varphi_m,$$
$$(PTC)\varphi_m = (-1)^m PT\varphi_m = (-1)^m\varphi_m = C\varphi_m = CPT\varphi_m.$$

Since $\mathrm{span}\{\psi_n\}_{n\in\mathbf{N}} = M$, we have $C^2 = I_M$, $HC = CH$, and $PT\cdot C = C\cdot PT$. Since $\forall \psi \in M$, we compute that

$$\langle PC\psi,\psi\rangle = \int_{\mathbf{R}} (TPC\psi)(x)\cdot\psi(x)dx$$

$$= \int_{\mathbf{R}} TP\left(\sum_n \varphi_n(x)\int_{\mathbf{R}} \varphi_n(y)\psi(y)dy\right)\cdot\psi(x)dx$$

$$= \int_{\mathbf{R}} \sum_n \left(\int_{\mathbf{R}} \varphi_n(y)\psi(y)dy\right)^* TP\varphi_n(x)\,\psi(x)dx$$

$$= \sum_n \left(\int_{\mathbf{R}} \varphi_n(y)\psi(y)dy\right)^* \int_{\mathbf{R}} \varphi_n(x)\,\psi(x)dx$$

$$= \sum_n \left|\int_{\mathbf{R}} \varphi_n(x)\,\psi(x)dx\right|^2$$

$$\geq 0.$$

Thus, we conclude that $PC$ is positive. Let $\langle PC\psi,\psi\rangle = 0$. Then for every $n$, we obtain that

$$0 = \int_{\mathbf{R}} \varphi_n(x)\,\psi(x)dx = \int_{\mathbf{R}} \varphi_n(-x)\,\psi(-x)dx = \int_{\mathbf{R}} (\varphi_n(x))^*\,\psi(-x)dx = \langle \varphi_n, P\psi\rangle.$$

It follows from $\text{span}\{\psi_n\}_{n\in\mathbf{N}} = M$ that $P\psi = 0$, i.e., $\psi = 0$. This shows that $\{C,P,T\}$ is *CPT* for $H$. Moreover, since $PT\varphi_n = \varphi_n$, $C\varphi_n = (-1)^n \varphi_n$, we have

$$\langle \varphi_m, \varphi_n\rangle_{CPT} = \langle PC\varphi_m, \varphi_n\rangle_{L^2}$$

$$= \int_{\mathbf{R}} (TPC\varphi_m)(x)\varphi_n(x)dx$$

$$= \int_{\mathbf{R}} (CTP\varphi_m)(x)\varphi_n(x)dx$$

$$= \int_{\mathbf{R}} (C\varphi_m)(x)\varphi_n(x)dx$$

$$= \int_{\mathbf{R}} (-1)^m \varphi_m(x)\varphi_n(x)dx$$

$$= \delta_{m,n}.$$

So, the eigenstates $\{\varphi_n\}$ are *CPT*-orthonormal. This completes the proof.

In the finite dimensional case, we have the following.

**Theorem 3.4.** *Let $\{P,T\}$ be a PT-frame on a finite dimensional Hilbert space $(M,\langle\cdot,\cdot\rangle)$, $H$ a linear operator on $M$ having unbroken PT-symmetry. Then $H$ has eigenstates $\varphi_n(n=1,2,\cdots,d)$ where $d = \dim M$ satisfying $PT\varphi_n = \varphi_n, H\varphi_n = E_n\varphi_n\ (n=1,2,\cdots,d)$. If in addition, $\text{span}\{\varphi_n : n=1,2,\cdots,d\} = M$, and*

$$C\left(\sum_{n=1}^d c_n\varphi_n\right) = \sum_{n=1}^d (-1)^n c_n\varphi_n,\quad \forall(c_1,c_2,\ldots,c_d)\in \mathbf{C}^d, \tag{19}$$

*then $\{C,P,T\}$ is a CPT-frame for $H$ satisfying $\langle\varphi_m,\varphi_n\rangle_{CPT} = \delta_{m,n}\ (m,n=1,2,\cdots,d)$ if and only if*

$$(\varphi_m,\varphi_n) := \langle P\varphi_m,\varphi_n\rangle_M = (-1)^m \delta_{m,n}(m,n=1,2,\cdots,d). \tag{20}$$

**Proof.** Similar to the beginning of the proof of Theorem 3.3, we see that $H$ has eigenstates $\varphi_n(n=1,2,\cdots,d)$ satisfying $PT\varphi_n = \varphi_n$, $H\varphi_n = E_n\varphi_n\ (n=1,2,\cdots,d)$. Next, we assume that $\text{span}\{\varphi_n : n=1,2,\cdots,d\} = M$. Then the formula defines a linear operator $C$ on $M$ with $C\varphi_n = (-1)^n \varphi_n\ (n=1,2,\cdots,d)$.

Suppose that $\{C,P,T\}$ is a *CPT*-frame for $H$ satisfying $\langle\varphi_m,\varphi_n\rangle_{CPT} = \delta_{m,n}$. Then for all $m,n = 1,2,\cdots,d$, we have

$$\langle P\varphi_m,\varphi_n\rangle_M = \langle P(-1)^m C\varphi_m,\varphi_n\rangle_M = (-1)^m\langle PC\varphi_m,\varphi_n\rangle_M = (-1)^m\langle \varphi_m,\varphi_n\rangle_{CPT} = (-1)^m\delta_{m,n}.$$

Conversely, we suppose that (20) holds. Since $\text{span}\{\varphi_n:n=1,2,\cdots,d\}=M$ and

$$PT\varphi_n = \varphi_n,\ H\varphi_n = E_n\varphi_n,\ C\varphi_n = (-1)^n\varphi_n\ (n=1,2,\cdots,d),$$

we get that $HC=CH$, $PTC=CPT$ and $C^2=I_M$. For every $x=\sum_{n=1}^{d}x_n\varphi_n\in M$, Eqn. (20) implies that

$$\langle PCx,x\rangle_M = \sum_{m,n=1}^{d}x_n^*x_m(-1)^n\langle P\varphi_n,\varphi_m\rangle_M = \sum_{m,n=1}^{d}x_n^*x_m(-1)^n(-1)^m\delta_{m,n} = \sum_{n=1}^{d}|x_n|^2\geq 0.$$

So, $PC$ is a positive definite operator on $M$. Hence, $\{C,P,T\}$ is a $CPT$-frame for $H$. Moreover,

$$(-1)^m\langle\varphi_m,\varphi_n\rangle_{CPT} = (-1)^m\langle PC\varphi_m,\varphi_n\rangle_M = \langle P\varphi_m,\varphi_n\rangle_M = (-1)^m\delta_{m,n}.$$

Thus, $\langle\varphi_m,\varphi_n\rangle_{CPT}=\delta_{m,n}$. The proof is completed.

**Remark 3.1.** Under conditions of Theorem 3.4, if the operator $H$ is symmetric, i.e., $H^T:=TH^+T=H$, then from the reality of $E_n$ we get that

$$E_m(\varphi_m,\varphi_n) = \langle PE_m\varphi_m,\varphi_n\rangle_M = \langle PH\varphi_m,\varphi_n\rangle_M$$
$$= \langle H^+P\varphi_m,\varphi_n\rangle_M = \langle P\varphi_m,H\varphi_n\rangle_M = E_n\langle P\varphi_m,\varphi_n\rangle_M = E_n(\varphi_m,\varphi_n).$$

Hence, $(\varphi_m,\varphi_n)=0$ whenever $E_m\neq E_n$.

Next example is an application of Theorem 3.4.

**Example 3.1.** Let $H,T,P$ be as in Example 2.1 with $s^2>r^2\sin^2\theta$. Then $H$ has eigenvalues $E_\pm = r\cos\theta\pm s\cos\varphi$ with corresponding eigenstates

$$|\psi_+\rangle = \frac{1}{\sqrt{2\cos\varphi}}\begin{pmatrix}e^{i\varphi/2}\\ e^{-i\varphi/2}\end{pmatrix},|\psi_-\rangle = \frac{1}{\sqrt{2\cos\varphi}}\begin{pmatrix}e^{-i\varphi/2}\\ -e^{i\varphi/2}\end{pmatrix}.$$

Clearly, $PT|\psi_\pm\rangle=\pm|\psi_\pm\rangle$ and so $H$ has unbroken $PT$-symmetry. Obviously,

$$(|\psi_\pm\rangle)^T|\psi_\pm\rangle = \frac{1}{2\cos\varphi}\left(e^{\pm i\varphi}+e^{\mp i\varphi}\right)=1, (|\psi_\pm\rangle)^T|\psi_\mp\rangle=0.$$

Put $|\varphi_-\rangle=-i|\psi_-\rangle, |\varphi_+\rangle=|\psi_+\rangle$, then $\{|\varphi_-\rangle,|\varphi_+\rangle\}$ is a basis for $\mathbf{C}^2$ satisfying

$$PT|\varphi_\pm\rangle=|\varphi_\pm\rangle, H|\varphi_\pm\rangle=E_\pm|\varphi_\pm\rangle, \langle P|\varphi_\pm\rangle,|\varphi_\pm\rangle\rangle = (TP|\varphi_\pm\rangle)^T|\varphi_\pm\rangle = (|\varphi_\pm\rangle)^T|\varphi_\pm\rangle=\pm 1.$$

It follows from Theorem 3.4 that the operator:

$$C:\mathbf{C}^2\to\mathbf{C}^2,\ C(a_-|\varphi_-\rangle+a_+|\varphi_+\rangle)=-a_-|\varphi_-\rangle+a_+|\varphi_+\rangle,$$

gives a $CPT$-frame $\{C,P,T\}$ for $H$ such that $\{|\varphi_-\rangle,|\varphi_+\rangle\}$ is an orthonormal basis for the Hilbert space $(\mathbf{C}^2,\langle\cdot,\cdot\rangle_{CPT})$. Since $C|\varphi_\pm\rangle=\pm|\varphi_\pm\rangle$, we see that

$$C = -|\varphi_-\rangle\langle\varphi_-|_{CPT} + |\varphi_+\rangle\langle\varphi_+|_{CPT} = |\varphi_-\rangle|\varphi_-\rangle^t + |\varphi_+\rangle|\varphi_+\rangle^t,$$

where $\langle\varphi_\pm|_{CPT} = \langle\varphi_\pm|PC = (CPT|\varphi_\pm\rangle)^T = \pm(|\varphi_\pm\rangle)^T$. An easy computation shows that

$$C = \begin{pmatrix}i\tan\varphi & \sec\varphi\\ \sec\varphi & -i\tan\varphi\end{pmatrix} = \frac{1}{\cos\varphi}\begin{pmatrix}i\sin\varphi & 1\\ 1 & -i\sin\varphi\end{pmatrix}. \qquad (21)$$

This is just the C operator obtained in [9]. Lastly, since $H$ is both symmetric and $PT$-symmetric, Corollary 3.2(6) implies that $H$ is $CPT$-Hermitian: $H^{CPT}=H$ and similar to the Hermitian matrix $h=(PC)^{1/2}H(PC)^{-1/2}$. Moreover, under basis $\{|\varphi_-\rangle,|\varphi_-\rangle\}$, the operators $H, C$ have the following matrix representations: $\begin{pmatrix} E_- & 0 \\ 0 & E_+ \end{pmatrix}, \begin{pmatrix} -1 & 0 \\ 0 & 1 \end{pmatrix}.$

The following can be checked easily.

**Theorem 3.4.** *Let* $\{C_i, P_i, T_i\}$ *be a* $C_i P_i T_i$-*frame on Hilbert space* $K_i$ *for* $i = 1, 2$, *and*

$$C = C_1 \otimes C_2, P = P_1 \otimes P_2, T = T_1 \otimes T_2, K = K_1 \otimes K_2.$$

*Then*

*(1)* $\{C, P, T\}$ *is a CPT-frame on the Hilbert space* $K$.

*(2) If in addition,* $\{C_i, P_i, T_i\}$ *(* $i = 1, 2$ *) is a* $C_i P_i T_i$-*frame for operator* $H_i$ *on* $K_i$, *respectively, then* $\{C, P, T\}$ *is a CPT-frame for the operator* $H := H_1 \otimes H_2$ *on* $K$.

*(3) If* $H_i$ *is* $P_i T_i$-*symmetric* $(i = 1, 2)$, *then* $H := H_1 \otimes H_2$ *is PT-symmetric.*

**Example 3.2.** Let

$$H_1 = \begin{pmatrix} r_1 e^{i\theta_1} & s_1 \\ s_1 & r_1 e^{-i\theta_1} \end{pmatrix}, H_2 = \begin{pmatrix} r_2 e^{i\theta_2} & s_2 \\ s_2 & r_2 e^{-i\theta_2} \end{pmatrix},$$

$$P_1 = P_2 = \begin{pmatrix} 0 & 1 \\ 1 & 0 \end{pmatrix}, T_k \begin{pmatrix} x \\ y \end{pmatrix} = \begin{pmatrix} \bar{x} \\ \bar{y} \end{pmatrix} (k = 1, 2),$$

$$C_1 = \begin{pmatrix} i \tan \varphi_1 & \sec \varphi_1 \\ \sec \varphi_1 & -i \tan \varphi_1 \end{pmatrix}, C_2 = \begin{pmatrix} i \tan \varphi_2 & \sec \varphi_2 \\ \sec \varphi_2 & -i \tan \varphi_2 \end{pmatrix},$$

where $r_k, s_k, \theta_k$ are nonzero real numbers satisfying $\sin \varphi_k = r_k / s_k \cdot \sin \theta_k$ ($|\varphi_k| < \pi/2$).
Define the following operators on $\mathbf{C}^2 \otimes \mathbf{C}^2$:

$$H = H_1 \otimes H_2, P = P_1 \otimes P_2, T = T_1 \otimes T_2, C = C_1 \otimes C_2.$$

From Example 2.1 and Theorem 2.5, we see that $\{C, P, T\}$ is a CPT-frame for $H$. Moreover, $H$ is PT-symmetric. Since $H$ has four eigenvalues:

$$E_+^1 E_+^2, E_+^1 E_-^2, E_-^1 E_+^2, E_-^1 E_-^2 \ (E_\pm^k \text{ as in Example 2.3}),$$

the corresponding eigenstates are as follows:

$$\left|\psi_+^1\right\rangle \otimes \left|\psi_+^2\right\rangle, \left|\psi_+^1\right\rangle \otimes \left|\psi_-^2\right\rangle, \left|\psi_-^1\right\rangle \otimes \left|\psi_+^2\right\rangle, \left|\psi_-^1\right\rangle \otimes \left|\psi_-^2\right\rangle,$$

($\left|\psi_\pm^k\right\rangle$ as in Example 2.3) which are the eigenstates of $PT$ for eigenvalues $1, -1, -1, 1$. Hence, $H$ has unbroken $PT$-symmetry. Clearly, $H$ is symmetric, and so $CPT$-Hermitian (Corollary 3.2(6)).

**Acknowledgements** This subject was supported by the National Natural Science Founds of China (10571113, 10871124, 11171197), the Fundamental Research Founds for the Central Universities (GK201002006) and the Innovation Founds for Graduate Program of Shaaxi Normal University (2011CXB004).

## References

[1] Bender C M, Boettcher S 1998 Real spectra in non-Hermitian Hamiltonians having *PT*-symmetry *Phys. Rev. Lett.* **80** 5243-5246

[2] Bender C M, Brody D C 2002 Complex extension of quantum mechanics *Phys. Rev. Lett.* **89** 270401-270404

[3] Bender C M, Brody D C, Jones H F, et al. 2007 Faster than Hermitian quantum mechanics *Phys. Rev. Lett.* **98** 040403-040406

[4] Bender C M, Klevansky S P 2010 Families of particles with different masses in PT-symmetric quantum field theory *Phys. Rev. Lett.* **105** 031601-031604

[5] Guo A, Salamo G J 2009 Observation of *PT* - symmetric breaking in complex optical

potentials *Phys. Rev. Lett.* **103** 093902-093905
[6] Ruter C E, Makris K G 2010 Observation of Parity-time symmetric in optical *Nature Physics* **192** 6
[7] Bender C M, Jones H F 2005 Dual *PT*-symmetric quantum field theories *Phys. Lett. B* **625** 333-340
[8] Bender C M, Chen J H, Milton K A 2006 *PT*-symmetric versus Hermitian formulations of quantum mechanics *J. Phys. A* **39** 1657-1668
[9] Bender C M 2007 Making sense of non-Hermitian Hamiltonians *Rep. Prog. Phys.* **70** 947-1018
[10] Bender C M, Hook D W 2008 Conjecture on the analyticity of $PT$-symmetric potentials and the reality of their spectra *J. Phys. A* **41** 392005-392014
[11] Bender C M, Brody D C, Hook D W 2008 Quantum effects in classical systems having complex energy *J. Phys. A* **41** 352003-352018
[12] Bender C M, Hook D W 2008 Exact isospectral pairs of *PT*-symmetric Hamiltonians *J. Phys. A* **41** 244005-244022
[13] Bender C M, Hook D W, Meisinger P N, et al. 2010 Probability density in the complex plane *Ann. Phys.* **325** 2332–2362
[14] Bender C M, Mannheim P D 2010 *PT*-Symmetry and necessary and sufficient conditions for the reality of energy eigenvalues *Phys. Lett. A* **374** 1616-1620
[15] Moiseyev N 2011 Non-Hermitian quantum mechanics (Cambridge University Press)
[16] Cho M H, Wu J D 2012 *PT*-Symmetry Scientific *J. Math. Res.* **2** 1-6
[17] Chong Y D, Ge L and Stone A D 2011 PT-symmetry breaking and laser-absorber modes in optical scattering systems *Phys. Rev. Lett.* **106** 093902
[18] Lin Z, Ramezani H, Eichelkraut T, et al. 2011 Unidirectional invisibility induced by *PT*-symmetric periodic structures *Phys. Rev. Lett.* **106** 213901
[19] Mannheim P D and O'Brien J G 2011 Impact of a global quadratic potential on galactic rotation curves *Phys. Rev. Lett.* **106** 121101
[20] Feng L, Ayache M, Huang J, et al. 2011 Nonreciprocal light propagation in a silicon photonic circuit *Science* **333** 729
[21] Bittner S, Dietz B, Guenther U, et al. 2012 *PT* symmetry and spontaneous symmetry breaking in a microwave billiard *Phys. Rev. Lett.* **108** 024101
[22] Liertzer M, Ge Li, Cerjan A, et al. 2012 Pump-induced exceptional points in lasers above threshold *Phys. Rev. Lett.* **108** 173901
[23] Zezyulin A, Konotop V V 2012 Nonlinear modes in finite-dimensional *PT*-symmetric systems *Phys. Rev. Lett.* **108** 213906
[24] Ramezani H, Christodoulides D N, Kovanis V, et al. 2012 *PT*-symmetric Talbot effects *Phys. Rev. Lett.* **109** 033902
[25] Regensberger A, Bersch C, Miri M A, et al. 2012 Parity-time synthetic photonic lattices *Nature* **488** 167